\documentclass[aps,prl,reprint,preprintnumbers,superscriptaddress,
nobibnotes,nofootinbib,longbibliography,amsmath,amssymb]{revtex4-2}

\usepackage{graphicx}
\usepackage{dcolumn}
\usepackage{bm}
\usepackage{color}
\usepackage{ulem}
\usepackage{gensymb}
\usepackage{braket}
\usepackage{amsmath}
\usepackage[percent]{overpic}
\usepackage{hyperref}
\DeclareUnicodeCharacter{221A}{\ensuremath{\sqrt{}}}

\begin{document}

\title{Recent progress on liquid transport growth of quantum materials}

\author{J.-Q. Yan}
\email{yanj@ornl.gov}
\affiliation{Materials Science and Technology Division, Oak Ridge National Laboratory, Oak Ridge, Tennessee 37831, USA}

\author{B. R. Ortiz}
\affiliation{Materials Science and Technology Division, Oak Ridge National Laboratory, Oak Ridge, Tennessee 37831, USA}

\date{\today}

\begin{abstract}

Liquid transport growth (LTG) is a horizontal flux growth technique that is closely analogous to chemical vapor transport, with the key distinction that a molten flux rather than a vapor serves as the transport agent. Unlike conventional flux growth, LTG spatially separates charge dissolution and crystal precipitation and couples them through continuous solute transport under a deliberately imposed temperature gradient. This enables crystal growth to begin before the charge is completely dissolved, removes the equilibrium solubility constraint on the starting charge/flux ratio, and allows large yields of single crystals to be obtained from a single growth. Recent studies have further shown that by spatially separating dissolution and crystallization and maintaining crystallization at a nearly constant temperature, LTG is particularly effective for two classes of materials: compounds that crystallize only within a narrow temperature and/or composition window, and compounds whose stoichiometry, defect concentration, and thus physical properties are sensitive to the crystallization temperature. In this review, we discuss representative examples including Fe$_3$Sn$_2$, CrTe$_3$, YFe$_2$Ge$_2$, UTe$_2$, CeRh$_2$As$_2$, MoTe$_2$, WTe$_2$, and LuNb$_6$Sn$_6$ to illustrate the unique capabilities of LTG for producing high quality single crystals of diverse quantum materials. We also summarize practical considerations for LTG experimental design, including furnace selection, growth time, melt stability, and ampoule geometry, and discuss future opportunities for transforming LTG from an empirical growth method into a more predictive crystal growth technique.

\end{abstract}

\maketitle

\subsection{Introduction}

Flux growth has long been one of the most powerful techniques for producing high quality single crystals of quantum materials \cite{Canfield1992MetallicFlux, Fisk2001HighTemperatureSolution, Fisk1989FluxGrowth, Kanatzidis2005MetalFlux, Wanklyn1972FluxGrowth, Bugaris2012FluxOxides, Scheel1983PresentStatusFluxGrowth, Phelan2012AdventuresCrystalGrowth}. It has played a central role in enabling studies of intrinsic physical properties by providing crystals with low defect densities and has also facilitated the synthesis and discovery of many new materials. Conventionally, flux growth is performed in a vertical configuration, where supersaturation and subsequent crystal growth are driven by cooling a homogeneous melt or occasionally by evaporating the molten flux. In recent years, horizontal flux growth has emerged as a useful complement to conventional vertical flux growth. In this approach, the charge continuously dissolves at the higher-solubility hot end, and the dissolved species are transported to the lower-solubility cold end, where supersaturation drives crystal growth. This spatial separation can overcome several limitations of conventional flux growth and enable growth in challenging material systems.

\begin{figure*} \centering \includegraphics [width = 0.6\textwidth] {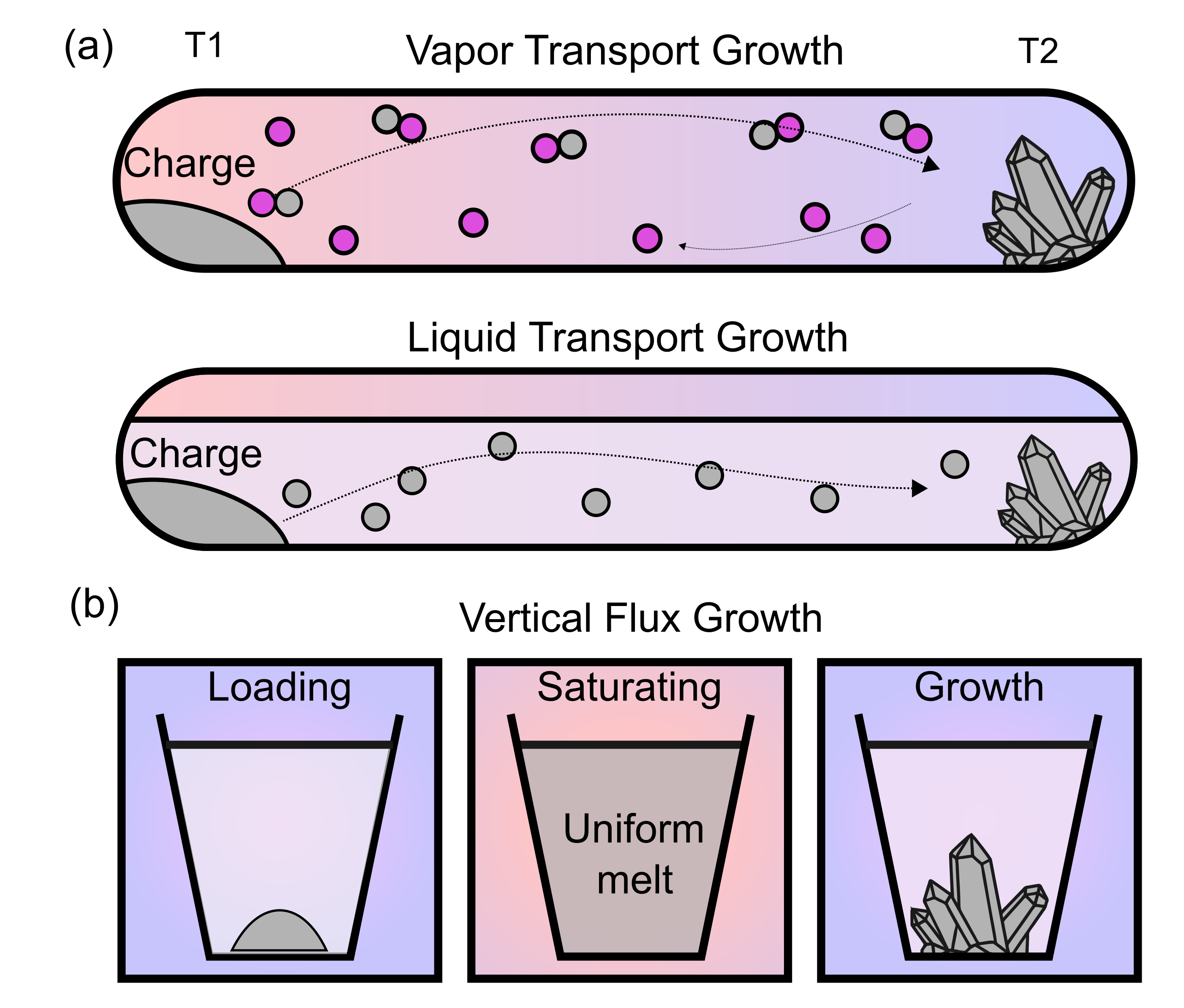}
\caption{(color online) Schematic comparison of chemical vapor transport (CVT), liquid transport growth (LTG), and conventional vertical flux growth. In both CVT and LTG, the source material is transported from a hot region to a colder crystallization region under a temperature gradient. The key distinction is that the transport medium is a vapor in CVT and a molten flux in LTG. In conventional vertical flux growth, crystallization occurs from a homogeneous melt during cooling.}
\label{Comp-1}
\end{figure*}

Related transport growth in molten halide flux under a temperature gradient was reported in the late 1960s and early 1970s for cubic ZnS and CuCl crystals \cite{Parker1968ZnS,Parker1970CuCl}. This approach received little attention until renewed interest emerged in the 2010s. Chareev et al. demonstrated the growth of FeSe single crystals from KCl-AlCl$_3$ flux and subsequently established general principles for growing chalcogenide and pnictide crystals in alkali metal and aluminum halide melts using a steady temperature gradient \cite{chareev2013single, chareev2016general}. Flux growth in horizontal configuration was also established in metallic flux systems motivating the broader concept of liquid transport growth (LTG) \cite{yan2017flux}.  LTG refers to transport driven solution growth sustained by continuous solute transport under a deliberately imposed temperature gradient, in close analogy to chemical vapor transport but using a molten flux rather than a vapor as the transport agent. This mechanistic similarity motivated us to introduce the term LTG for this approach. Figure~\ref{Comp-1} compares the growth mechanisms of chemical vapor transport, LTG, and conventional vertical flux growth. In a typical LTG experiment, the charge is placed at the hot end of the growth ampoule, where it continuously dissolves into the molten flux. The imposed temperature gradient establishes a gradient in equilibrium solubility, and the dissolved solute is transported toward the cold end where the lower solubility generates supersaturation and drives crystal nucleation and growth. In contrast, conventional vertical flux growth doen't deliberately separate the dissolution and crystallization zones and typically generates supersaturation by cooling the bulk solution or evaporating the molten flux. As summarized in Table~\ref{tab:LTG}, LTG is distinguished by three key characteristics: (1) the spatial separation of charge dissolution and crystal precipitation, (2) independent temperatures for charge dissolution and crystallization, and (3) crystallization at a nearly constant temperature. These characteristics lead to several practical advantages over conventional vertical flux growth, including large crystal yields that are no longer limited by the equilibrium solubility of the charge in the flux, independent optimization and control of charge dissolution and crystallization conditions, and improved crystal homogeneity for compounds with temperature sensitive nonstoichiometry and/or lattice defects.

\begin{table*}[t]
\caption{Fundamental design principles of liquid transport growth, their physical consequences, situations where LTG should be considered, and representative examples discussed in this review.}
\label{tab:LTG}

\centering
\renewcommand{\arraystretch}{1.4}
\setlength{\tabcolsep}{10pt}      

\begin{tabular}{p{3.6cm} p{5.5cm} p{3.8cm} p{3.7cm}}
\hline\hline

\textbf{Key characteristic of LTG} &
\textbf{Physical consequence} &
\textbf{When to consider LTG} &
\textbf{Representative examples} \\

\hline

Charge dissolution and crystallization are spatially separated

& - Crystal growth begins before the charge is completely dissolved.

- Continuous transport of dissolved species to the growth front allows the yield of a large quantity of crystals and sometimes large crystals.

& - One or more constituents have limited solubility in the flux. 

- A large quantity of crystals is required.

-The temperature required for efficient charge dissolution differs from that needed to achieve the desired stoichiometry, defect concentration, or phase stability during crystallization.

& Mo$_3$Sb$_7$\cite{yan2017flux}, IrSb$_3$\cite{yan2017flux}, FeSn, Fe$_3$Sn$_2$\cite{wang2021stimulated}, Fe$_5$Sn$_3$, FeSe\cite{yan2017flux,chareev2013single}, LuNb$_6$Sn$_6$, ScV$_6$Sn$_6$, MoTe$_2$\cite{Park2026FCI}, WTe$_2$\cite{Delgado2026WTe2}  \\

\hline

Charge dissolution and crystallization occur at different temperatures

& - The hot end can be optimized for dissolution while the cold end for crystallization. 

- The temperature profile along the ampoule can be deliberately engineered to control supersaturation and growth kinetics.

& - The desired phase is stable only within a narrow temperature and/or composition window.

- Independent optimization of charge dissolution and crystallization is needed to maintain melt stability.

- Precise control of chemical stoichiometry and defect concentration is needed.

& Fe$_3$Sn$_2$\cite{wang2021stimulated},  NiPSe$_3$\cite{yan2017flux}, CrTe$_3$, Cr$_{11}$Ge$_{19}$, CeRh$_2$As$_2$\cite{chajewski2024horizontal}  \\

\hline

Crystallization occurs at a nearly constant temperature

& - Nearly constant growth conditions promote uniform stoichiometry and defect populations throughout the crystal.

& - The target material's nonstoichiometry, defect concentration, and resulting its physical properties are sensitive to the crystal growth temperature.
- Precise control of dopant concentration is required in systems whose dopant distribution coefficients are sensitive to the growth temperature.

& UTe$_2$\cite{aoki2024molten}, MoTe$_2$\cite{Park2026FCI}, WTe$_2$\cite{Delgado2026WTe2}, YFe$_2$Ge$_2$\cite{chen2020unconventional}, CeRh$_2$As$_2$\cite{chajewski2024horizontal},   LuNb$_6$Sn$_6$ \\

\hline\hline

\end{tabular}
\end{table*}

The ability to obtain high yields of single crystals is the most immediate practical advantage of LTG. Because the charge dissolution continuously supplies solute to the growth front located in a spatially separated region, the total charge loading can greatly exceed the amount that is soluble in the flux at any one time. The attainable yield is therefore not directly limited by the equilibrium solubility of the charge in the flux, as highlighted by the growth of Mo$_3$Sb$_7$ in our initial LTG report \cite{yan2017flux}. In addition to a high yield, the spatial separation of the source and growth regions partially decouples the dissolution and crystallization conditions, providing greater freedom to optimize each process. The source region can be maintained under conditions favorable for charge dissolution while the growth region is held within the narrow temperature and/or composition window required for crystallization of the target phase, as illustrated by the growths of Fe$_3$Sn$_2$ \cite{wang2021stimulated} and CrTe$_3$ reported in this work. Moreover, because crystallization occurs at a nearly constant temperature, LTG avoids the continuously changing phase equilibrium and defect chemistry associated with cooling the entire melt. This feature can improve compositional and defect homogeneity in materials whose nonstoichiometry, defect concentrations, and physical properties are highly temperature-sensitive, as demonstrated by YFe$_2$Ge$_2$ \cite{chen2020unconventional}, UTe$_2$ \cite{aoki2024molten}, CeRh$_2$As$_2$ \cite{chajewski2024horizontal}, MoTe$_2$ \cite{Park2026FCI}, and WTe$_2$ \cite{Delgado2026WTe2}.

In recent years, applications of LTG span transition-metal chalcogenides, halides, pnictides, thiophosphates, selenophosphates, and oxides from halide salt fluxes \cite{yan2017flux,mcguire2017antiferromagnetism,mcguire2019chemical,chareev2013single,chareev2016synthesis,chareev2016general,chareev2017single,chareev2018single,sun2023coexistence,aoki2024molten,ma2023growth,wang2024structural,ruixian2024low,kothakonda2023high,zhu2021synthesis,mandal2021superconductivity,cevallos2019liquid,xing2021synthesis,Stilkerich2026}, ultimate sulfides from sulfur flux \cite{Chareev2025UltimateSulfides}, and a wide variety of compounds from metallic fluxes. Representative self flux examples include IrSb$_3$, Mo$_3$Sb$_7$, and MnBi \cite{yan2017flux}, Fe$3$Sn$2$ \cite{wang2021stimulated}, LaSbTe \cite{Bannies2026LaSbTe}, Ir$_{3.8}$Sb$_{12}$ \cite{Wang2025Ir3p8Sb12}, LuNb$_6$Sn$_6$ and MoTe$_2$ \cite{Park2026FCI}, WTe$_2$ \cite{Delgado2026WTe2}, and IrSn$_4$ \cite{nakamura2023fermi}. Other metallic-flux examples include CeRh$_2$As$_2$ grown from Bi flux \cite{chajewski2024horizontal}, YFe$_2$Ge$_2$ from Sn flux \cite{chen2020unconventional}, and $\alpha$-Mn from Pb flux \cite{manago2022site}.

In this review, we use representative applications from the published literature and experimental observations from our laboratory to illustrate the unique capabilities of LTG for producing high quality single crystals of quantum materials and to motivate its broader use in challenging materials systems. Drawing on these examples and the comparison summarized in Table~\ref{tab:LTG},  we provide practical guidance for determining when LTG is likely to offer advantages over conventional vertical flux growth for a given material system. We also discuss practical considerations for LTG experimental design and describe a dumbbell shaped ampoule that reduces flux consumption and facilitates crystal/flux separation. Finally, we consider opportunities to transform LTG from an empirical method into a more predictive crystal growth technique, and to explore its unique design flexibility to enable the growth and discovery of novel quantum materials.

\begin{figure*} \centering \includegraphics [width = \textwidth] {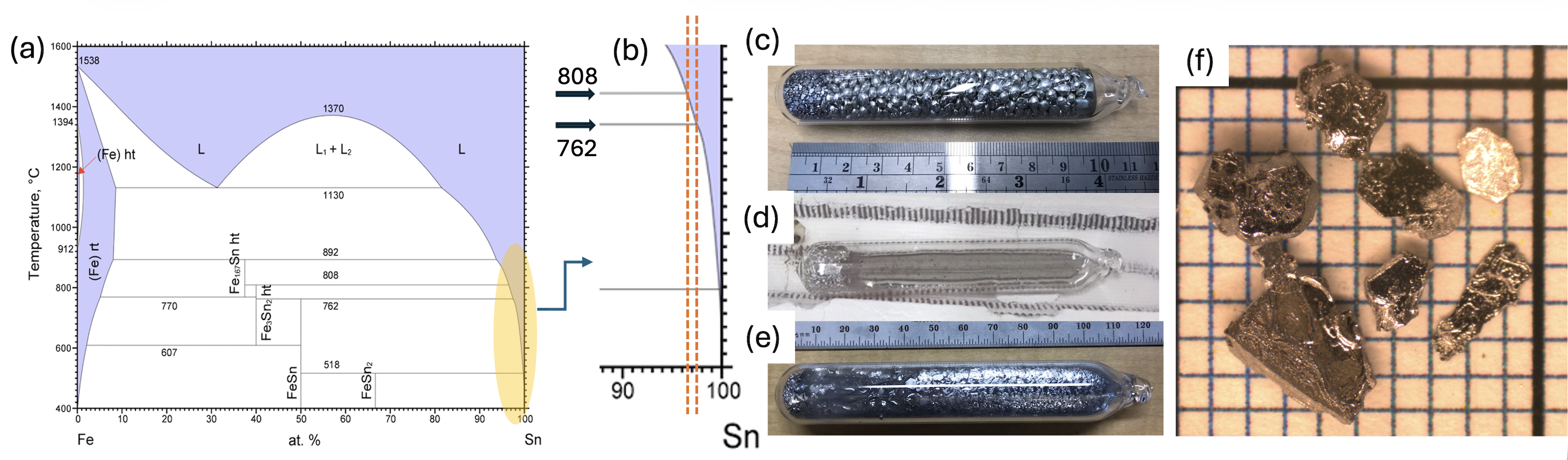}
\caption{(color online) Liquid transport growth of Fe$_3$Sn$_2$. (a) Phase diagram of Fe-Sn from Reference\,[\citenum{Okamoto2016}]. (b) Portion of the phase diagram highlighting the narrow composition (vertical dashed lines) and temperature (horizontal solid arrows) ranges for the conventional vertical flux growth. (c) The sealed growth ampoule  with Fe pieces staying at one end of the ampoule and not mixed with Sn flux. (d) Growth ampoule at 300$\degree$C after Sn flux melts. (e) The ampoule after growth. (f) Picture of Fe$_3$Sn$_2$ crystals on a millimeter grid.}
\label{FeSn-1}
\end{figure*}

\subsection{Narrow temperature and/or composition range}

Occasionally, the temperature and/or composition range suitable for flux growth of a specific phase can be rather narrow. In a conventional vertical flux growth, a narrow crystallization temperature range often results in small crystals, precipitation of neighboring phases, and difficulty separating crystals from the remaining flux by decanting. Likewise, when the composition window is narrow, the starting composition must be carefully chosen to avoid the formation of neighboring phases in phase diagram. In both situations, LTG offers significant advantages over conventional vertical flux growth. Since crystallization occurs first only at the cold end of the growth ampoule in LTG, only the temperature near the cold end needs to lie within the stability range of the desired phase. The hot end can be maintained at a substantially higher temperature to enhance charge dissolution and its transport through the molten flux without affecting the identity of the crystallizing phase. More importantly, the temperature at the hot end can be adjusted independently to optimize the dissolution rate, mass transport, and ultimately the growth kinetics. Fe$_3$Sn$_2$ and CrTe$_3$ are representative examples in which the temperature and composition windows are both narrow for self flux growth in the conventional vertical configuration. We present the details of LTG of these compounds performed in our lab to illustrate how LTG proves advantageous over conventional vertical flux growth for these cases.

\begin{figure*} \centering \includegraphics [width = 0.8\textwidth] {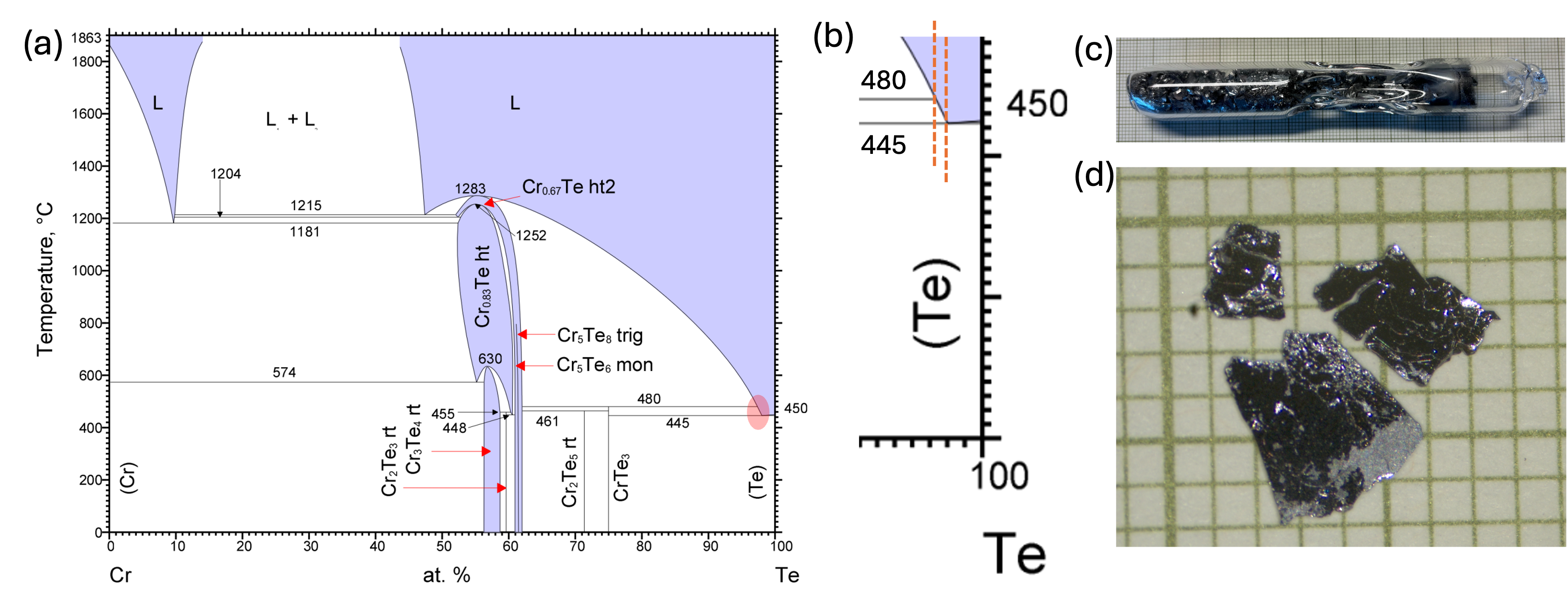}
\caption{(color online) Liquid transport growth of CrTe$_3$. (a) Cr-Te binary phase diagram from Reference\,[\citenum{Okamoto2016}]. The region covered by an oval shaped circle is highlighted in (b) to illustrate the composition and temperature range good for the conventional vertical flux growth. (c) The growth ampoule on a millimeter grid. Part of the quartz tube was collapsed using a torch in order to reduce the amount of flux needed for the growth. See the Section on growth ampoule for more details. (d) CrTe$_3$ crystals on a millimeter grid. }
\label{CrPD-1}
\end{figure*}

\textit{The case of Fe$_3$Sn$_2$:} Three Fe-Sn binary compounds FeSn, Fe$_3$Sn$_2$, and Fe$_5$Sn$_3$ all contain kagome arrangements of Fe atoms that give rise to flat bands in their electronic structures. According to the Fe–Sn binary phase diagram shown in Figure~\ref{FeSn-1}(a), each of these compounds is, in principle, accessible by conventional vertical flux growth from a Sn-rich melt. Indeed, most previous studies have grown these compounds using conventional vertical flux growth out of Sn melt. Among all three compounds, however, Fe$_3$Sn$_2$ is particularly challenging because it coexists with liquid only within a narrow temperature range from 762~$^\circ$C to 808~$^\circ$C. As highlighted in Figure\,\ref{FeSn-1}(b), the Fe concentration in the starting melt must be carefully controlled within a narrow range of approximately 2.5-3.5 at.\% to avoid the formation of neighboring FeSn and Fe$_5$Sn$_3$. The simultaneous constraints on temperature and composition make it difficult to obtain sizable single crystals without precipitating competing phases. Furthermore, because the starting melt contains only about 3 at.\% Fe, whereas Fe$_3$Sn$_2$ contains 60 at.\% Fe, the crystal yield from a conventional vertical flux growth is intrinsically limited.

Using LTG for this situation greatly simplifies the growth because the phase selection is controlled primarily by the temperature at the cold end of the ampoule. Figures\,\ref{FeSn-1}(c-e) show the details of the LTG of Fe$_3$Sn$_2$ performed in our laboratory. In a typical growth, 10 grams of Fe pieces (Alfa Aesar, 1$\sim$3 mm, 99.98\%) and 60 grams of Sn shot (Alfa Aesar, 99.9999\%) were sealed under vacuum in a 10\,cm long quartz tube of 19 mm outer diameter, 1.5 mm wall thickness \cite{wang2021stimulated}.  As shown in Figure\,\ref{FeSn-1}(c), all Fe pieces were placed at one end of the quartz tube and were not mixed with Sn flux. The growth ampule was then put in a single zone tube furnace with the Fe pieces located near the center of the tube furnace.  Figure\,\ref{FeSn-1}(d) shows the picture of the ampoule at 300$\degree$C after Sn flux had melted. A 20$\degree$C natural temperature gradient along the growth ampoule was employed during the crystal growth with the furnace set at 802$\degree$C and the cold end of the growth ampoule at 782$\degree$C. This temperature is obtained by putting some insulations near the end of the heating elements. After two weeks, the furnace was powered off. In general, a longer growth time leads to more  crystals pushing the growth front toward the hot end of the growth ampoule.  Figure\,\ref{FeSn-1}(e) shows a picture of the ampoule after the crystal growth. A considerable amount of Fe remains undissolved at the hot end but they do not disturb the growth at the cold end. To extract Fe$_3$Sn$_2$ crystals, a portion of the ingot near the cold end was dissolved in concentrated hydrochloric acid that dissolves Sn flux but not the crystals. This process can be time consuming when a significant amount of Sn flux remains trapped between the crystals. To expedite the crystal extraction, we  loaded a portion of the ingot into a Canfield crucible set \cite{canfield2016use}, sealed them inside of a quartz tube under vacuum, kept it at 790$\degree$C for 3 hours, and then centrifuged to remove most of the Sn flux. Figure\,\ref{FeSn-1}(f)  shows the picture of crystals obtained after decanting. Some residual Sn flux is normally observed on the surface of the crystals and can be further removed using hydrochloric acid as needed.

The Fe$_3$Sn$_2$ growth highlights several practical advantages of LTG. The initial charge/flux ratio can exceed the equilibrium solubility and the undissolved Fe pieces at the hot end serve as a continuous solute reservoir. Provided that the hot end is sufficiently warm to sustain charge dissolution and transport, the target phase is selected primarily by the crystallization temperature at the cold end. Continuous supply of the dissolved solute to the growth front then enables a large quantity of Fe$_3$Sn$_2$ crystals to be obtained from a single growth.

It is worth mentioning that we have used LTG to grow all three Fe-Sn binary compounds, FeSn, Fe$_3$Sn$_2$, and Fe$_5$Sn$_3$ in order to obtain large quantities of single crystals for neutron scattering measurements. For all three compounds, the Fe content in the starting melt appropriate for conventional vertical flux growth out of Sn is only about 1-6~at.\%, much lower than the Fe content in the corresponding crystalline phases. As a result, the crystal yield is intrinsically small in the conventional vertical flux growth. For example, in our previous vertical flux growth of FeSn crystals, 0.33\,g of Fe was mixed with 34\,g of Sn resulting in about 0.6\,gram of crystals in a typical vertical flux growth \cite{sales2019electronic}. This amount is much smaller than the quantity recoverable from only the coldest $\sim$1~cm portion of the ingot in a typical LTG ampoule. This comparison highlights the much higher crystal yield achievable in LTG, where the Fe source at the hot end continuously supplies material to the crystallization region at the cold end without requiring all of the Fe to dissolve in the Sn-rich melt at once.

It is also worth noting that, in the LTG growths of all three Fe–Sn binary compounds, we used quartz tubes of the same dimensions, the same amounts of Fe and Sn, and the same temperature difference of 20~$^\circ$C along the ampoule. The only parameter intentionally varied was the temperature at the cold end, which controlled which Fe-Sn phase crystallized. This demonstrates a practical advantage of LTG: once an appropriate transport condition is established, the desired phase can be selected primarily by controlling the crystallization temperature at the cold end, rather than by precisely tuning the starting composition of a homogeneous melt.

\textit{The case of CrTe$_3$:} Figure\,\ref{CrPD-1}(a) shows the Cr-Te binary phase diagram. As highlighted in Fig.\,\ref{CrPD-1}(b), the liquid coexisting with CrTe$_3$ has a very narrow composition range (97\%-98\% Te) and the self flux growth in vertical configuration should be performed in a very narrow temperature range (445-480$\degree$C). Despite the narrow composition and temperature ranges, conventional vertical flux growth is still possible although a careful control of the starting composition and growth temperature profile is necessary \cite{mcguire2017antiferromagnetism}. In such a growth, we put an external thermocouple right next to the growth ampoule to carefully monitor the growth temperature. In our test growths, we noticed the precipitation of Cr$_5$Te$_8$ when the Cr content in the starting materials is slightly over 3\%, highlighting the importance of a careful control of the starting composition. 

CrTe$_3$ can be grown out of KCl-AlCl$_3$ flux using LTG \cite{mcguire2017antiferromagnetism}. Here we provide details for LTG of CrTe$_3$ out of Te flux. Figure\,\ref{CrPD-1}(c) shows the growth ampoule on a millimeter grid. In this growth, about 3 grams of Cr granules were first loaded into the quartz ampoule, and then about 40\,g Te shots were added to fill the remaining space. The temperature at the cold end was maintained at 450$\degree$C and a temperature difference of 30$\degree$C was applied along the growth ampoule. After about three weeks, a large quantity of millimeter sized single crystals was obtained at the cold end of the growth ampoule. Fig.\,\ref{CrPD-1}(d) shows the picture of some typical crystals after decanting from flux. We varied the Cr mass from 3 to 6\,g while keeping similar amount of Te in our test growths and found that the initial Cr/Te ratio did not need to be precisely controlled. Once sufficient transport of Cr through the Te flux was established, maintaining the cold end temperature between 445 and 480~$\degree$C provided an effective window for the crystallization of CrTe$_3$. Together with the Fe-Sn examples presented above, the growth of CrTe$_3$ illustrates a key advantage of LTG for systems with narrow phase stability windows: the overall charge/flux ratio needs not be precisely controlled, while the desired phase can be selected primarily through local control of the crystallization temperature. For Fe$_3$Sn$_2$ and CrTe$_3$, the growth outcome was relatively insensitive to the initial charge/flux ratio within the ranges examined. This behavior is not universal, however, because rapid charge dissolution can make melt stability sensitive to the initial loading as discussed later in Experimental Considerations.

\begin{figure*} \centering \includegraphics [width = 0.8\textwidth] {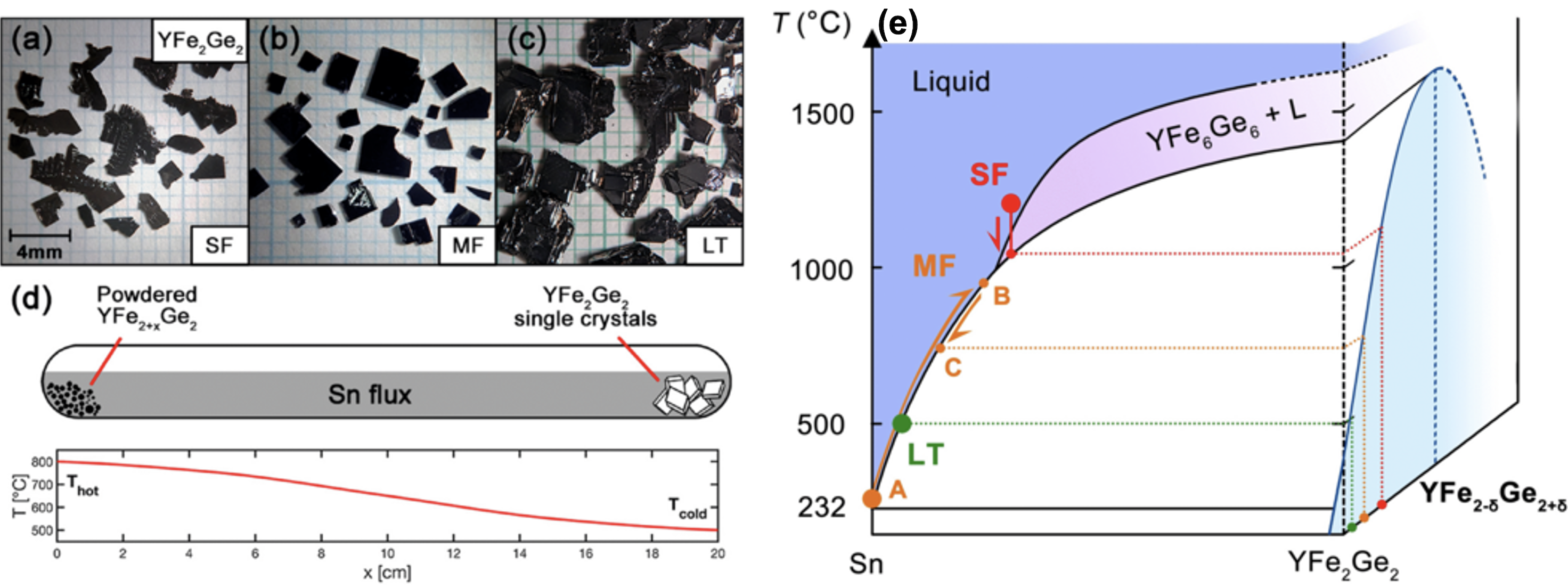}
\caption{(color online) Single crystal growth of YFe$_2$Ge$_2$. (a-c) Pictures of single crystals grown using (a) standard flux (SF), (b) modified flux (MF), and (c) liquid transport (LT) techniques. (d) Schematic illustration of the growth ampoule and temperature gradient used in the LTG. (e) Schematic phase diagram for the growths of  YFe$_2$Ge$_2$ single crystals out of Sn flux. Reproduced with permission from [\citenum{chen2020unconventional}].}
\label{YFe2Ge2-1}
\end{figure*}

\subsection{Temperature dependent nonstoichiometry and defects}

When the nonstoichiometry or defect concentration of a material is sensitive to the crystallization temperature, obtaining chemically uniform single crystals using the conventional vertical flux technique can be challenging. Cooling a homogeneous melt through a temperature range, a general approach to generating supersaturation in a vertical flux growth, can result in chemical inhomogeneity of the grown crystals. If the desired phase has a temperature dependent homogeneity range or defect population, different parts of a crystal, or crystals nucleated at different stages of the growth, may incorporate different vacancy concentrations, antisite disorder, dopant levels, or off-stoichiometry. Such growth induced inhomogeneity can sometimes significantly alter the physical properties and hinder the study of their intrinsic characteristics. A possible remedy is to perform the vertical flux growth at a fixed temperature provided that the crystal nonstoichiometry is less sensitive to changes in melt composition \cite{huang2016spectroscopic}. For fixed temperature growths performed in open crucibles, the evaporation of flux can generate the supersaturation necessary for the crystal growth. For growths performed in sealed containers, however, fixed temperature vertical flux growth is more difficult and crystals obtained are in general small due to the uncontrolled nucleation at the initial stage of the growth and lack of a temperature gradient inside of the crucible, irrespective of whether the growth begins with a uniform mixture or not. The starting composition affects the viscosity of the melt at the growth temperature, which can have a dramatic effect on how the solute atoms incorporate into the growing crystal lattice. 

LTG provides a straightforward way to address these limitations by separating charge dissolution, mass transport, and crystallization in space. Supersaturation at the cold end is generated by continuous dissolution of the charge at the hot end and transport of dissolved species through the molten flux, rather than by cooling the entire melt. As a result, crystallization can occur at a nearly constant temperature selected by the cold end of the ampoule. This fixed crystallization temperature can help maintain uniform stoichiometry and defect concentration throughout the growth. At the same time, the temperature profile along the ampoule can be adjusted independently to control dissolution, transport, supersaturation, and crystallization. In particular, the temperature variation near the crystallization zone can be deliberately minimized to reduce sample-to-sample or position-dependent variations. Therefore, a large quantity of sizable single crystals with improved chemical homogeneity can be obtained. Recent studies on YFe$_2$Ge$_2$ \cite{chen2020unconventional}, CeRh$_2$As$_2$ \citep{chajewski2024horizontal}, and UTe$_2$ \citep{aoki2024molten} exemplify this advantage and demonstrate that LTG can produce high-quality single crystals for compounds with growth temperature sensitive nonstoichiometry. Similar effects have also been observed in materials where the relevant defects or disorder are not yet fully understood. For example, LTG has produced MoTe$_2$ and WTe$_2$ crystals\cite{Park2026FCI, Delgado2026WTe2} with substantially reduced defect density and LuNb$_6$Sn$_6$ crystals with a density-wave transition at a higher temperature than crystals grown by conventional vertical flux growth. Although the microscopic origin of these improvements remains to be clarified, these results suggest that LTG is an effective and convenient technique for chemically diverse compounds whose stoichiometry, defect concentration, or physical properties are sensitive to the crystallization temperature. The constant temperature growth at reasonably low temperatures enabled by LTG may also be responsible for the high residual resistivity ratio (RRR) of IrSn$_4$ and Mn single crystals\cite{nakamura2023fermi, manago2022site} and the uniform dopant distribution in Fe$_{1-x}$Co$_x$Se single crystals\cite{wang2024structural}.

\textit{The case of YFe$_2$Ge$_2$:} YFe$_2$Ge$_2$ is a new unconventional superconductor with the well known ThCr$_2$Si$_2$ structure\cite{zou2014fermi}. Its superconducting properties are sensitive to lattice disorders such as Fe deficiency\cite{chen2019composition}. Chen et al. \cite{chen2020unconventional}  grew YFe$_2$Ge$_2$ single crystals out of Sn flux using three different methods: standard flux growth, modified flux growth, and LTG. A careful characterization of the nonstoichiometry, specific heat, magnetic and transport properties shows that LTG yields the best single crystals with a seven-fold reduction in disorder level. These ultra-pure single crystals allow for the observation of sharp thermodynamic transition anomalies indicative of bulk superconductivity and also the investigation of the intrinsic properties of this novel unconventional superconductor.

Figures\,\ref{YFe2Ge2-1}(a-c) show the YFe$_2$Ge$_2$ single crystals grown by three different flux methods out of Sn melt \cite{chen2020unconventional}. In the standard flux method, all starting elements Y, Fe, Ge, and Sn are sealed inside a quartz tube, homogenized at 1200$\degree$C and then cooled to 500$\degree$C in 75 hours.  The atomic ratio of the starting materials was varied  in order to prohibit the growth of YFe$_6$Ge$_6$ secondary phase and to increase the Fe content in the resulting YFe$_2$Ge$_2$ crystals. In the modified flux growth, polycrystalline YFe$_{2+x}$Ge$_2$ (0.02$<x<0.1$) produced in an induction furnace was used as the starting materials. The use of presynthesized polycrystalline materials lowers the peak temperature to between 850$\degree$C and 1150$\degree$C. The growth was performed by thermal cycling between the peak temperature and 500$\degree$C. In the LTG, the presynthesized polycrystalline starting material stays at the hot end at 800$\degree$C and crystal growth occurs at the cold end kept at 500$\degree$C.

Crystals obtained from these three different flux growths show distinct resistive, thermodynamic, and magnetic signatures of the superconducting transition.  Crystals from the standard flux method has the smallest RRR of 50-60, displays no zero resistivity or specific heat anomalies corresponding to the superconducting transition. Crystals from the modified flux method have a larger RRR near 200, show multiple resistivity drops before reaching zero and broad heat capacity anomalies as in polycrystalline samples. Crystals from LTG have the largest RRR of 430 and show a sharp heat capacity jump at a bulk superconducting transition temperature of about 1.1 K.

\begin{figure*} \centering \includegraphics [width = 0.8\textwidth] {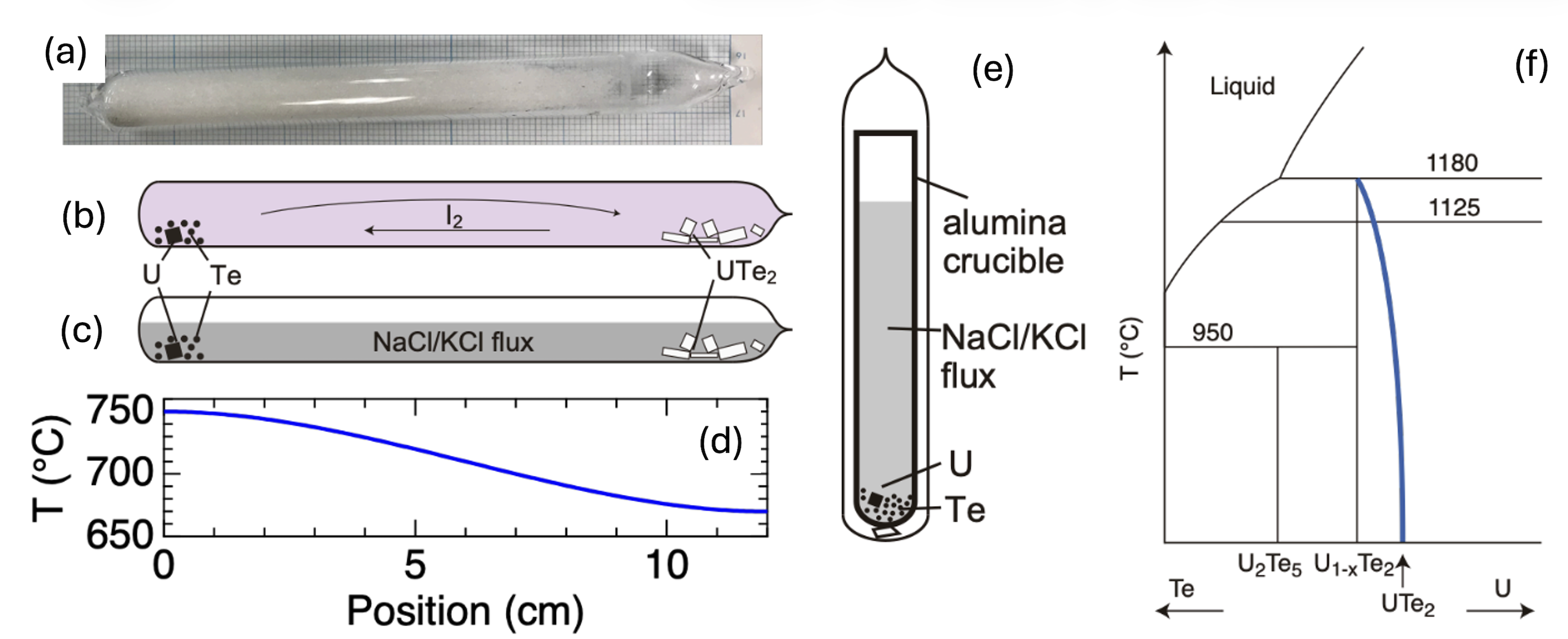}
\caption{(color online) Single crystal growth of UTe$_2$. (a) Ampoule for the liquid transport growth of UTe$_2$ out of NaCl/KCl flux. (b) Schematic illustration of the vapor transport growth of UTe$_2$ using I$_2$ as the transport agent. (c) Schematic illustration of the liquid transport growth of UTe$_2$ out of NaCl/KCl flux. (d) Temperature profile for the liquid transport growth. (e) Schematic illustration of the growth of UTe$_2$ out of NaCl/KCl flux using the conventional vertical configuration. (f) Schematic U-Te phase diagram with the blue curve showing possible temperature dependent nonstoichiometry of UTe$_2$. Reproduced with permission from [\citenum{aoki2024molten}].}
\label{UTe2-1}
\end{figure*}

The above difference in crystal quality can be understood using the phase diagram shown in Fig.\,\ref{YFe2Ge2-1}(e). YFe$_2$Ge$_2$ exhibits temperature sensitive nonstoichiometry and becomes richer in Ge at high temperatures. In both standard and modified flux growths,  crystallization occurs at relatively high temperatures and cooling over a wide temperature range leads to nonuniform crystal compositions. The improved crystal quality of crystals from the modified flux technique results from the reduced peak growth temperatures and thus narrower crystallization temperature range achieved by using presynthesized polycrystalline starting materials. In contrast, in LTG the dissolution and crystallization processes are spatially separated: the polycrystalline starting material is kept at the hot end at 800$\degree$C to provide a continuous supply of solute, while crystallization occurs at the cold end at a much lower and nearly constant temperature of 500$\degree$C. This separation avoids the need to fully dissolve all starting materials in Sn flux and removes the solubility constraint that would otherwise limit low-temperature growth. Growth at this lower and fixed crystallization temperature favors the formation of YFe$_2$Ge$_2$ crystals with stoichiometry closer to the ideal composition. In addition, maintaining a constant crystallization temperature throughout the growth promotes chemical homogeneity and minimizes Fe-Ge site disorder.

Although we have been emphasizing the unique ability of LTG to produce a large total yield, the crystals shown in Figures~4(a)–(c) suggest that it can also support the growth of comparatively large individual crystals. Together with the large LuNb$_6$Sn$_6$ crystals discussed later, these results indicate that LTG can provide a large quantity of sizable single crystals when the kinetics of the charge dissolution and transport, crystal nucleation and growth is appropriately controlled.

\textit{The case of UTe$_2$:} UTe$_2$ has attracted considerable attention because of the possible spin triplet nature of its superconductivity \cite{ran2019nearly,aoki2019unconventional}. The importance of single crystal quality in establishing its intrinsic superconducting properties became apparent when the high residual density of states observed in specific heat measurements of lower quality samples led to the proposal of partially gapped superconductivity. In contrast, higher quality samples exhibit a nearly vanishing residual density of states and a single superconducting transition at ambient pressure and zero magnetic field.

Dai Aoki carefully characterized UTe$_2$ single crystals grown by different growth techniques including chemical vapor transport, vertical flux growth out of Te, flux growth out of molten salt in both vertical and horizontal configurations (illustrated in Figs.\,\ref{UTe2-1}(a-e)) \cite{aoki2024molten}. While the flux growth out of molten salt produces crystals of better quality than chemical vapor transport and vertical flux growth out of Te melt, the difference between vertical and horizontal flux growths out of molten salt flux is of particular interest and highlights that LTG works well for compounds with a temperature dependent nonstoichiometry.

A mixture of NaCl and KCl with an equal molar ratio was selected as the flux in the growths out of molten salt. Figure\,\ref{UTe2-1}(e) shows the schematic picture of the ampoule for the vertical flux growth. The starting materials and flux with a molar ratio of U : Te : NaCl : KCl = 1 : 1.65 : 30 : 30 were placed in a long alumina crucible or directly into a quartz ampoule. After dehydrating under vacuum at about 200$\degree$C, the quartz ampoule was sealed under vacuum, heated to 950$\degree$C and stayed at this temperature for a day, then slowly cooled to 650$\degree$C. The furnace was powered off after dwelling at 650$\degree$C for a day. Small single crystals with RRR\,=\,50–220 and superconducting temperature  Tc\,=\,2.00–2.08\,K are found at the bottom of the alumina crucible or quartz tube.

Figures\,\ref{UTe2-1}(c,d) show the schematic of the growth ampoule and temperature profile for the LTG out of molten salt flux. U and Te starting materials were kept at the hot end at 750$\degree$C, while crystallization occurs at the cold end maintained at 670$\degree$C. These crystals are of high quality by showing a sharp single specific heat jump at 2.05\,K and a RRR of 800.

The comparison between vertical and horizontal molten-salt flux growths shows that LTG produces higher quality UTe$_2$ crystals even when the same salt mixture is used. The difference between these two techniques is proposed to result from the temperature dependent nonstoichiometry of UTe$_2$ as illustrated in Fig.\,\ref{UTe2-1}(f). UTe$_2$  tends to be U deficient at higher temperatures. In conventional vertical flux growth, crystallization occurs during cooling over a wide temperature range, which results in nonuniform nonstoichiometry in crystals and also the formation of competing phases such as U$7$Te${12}$ and U$_3$Te$_5$. In contrast, LTG maintains the crystallization region at a relatively low and nearly constant temperature throughout the growth. This low and nearly constant crystallization temperature likely minimizes growth temperature induced variations in U deficiency and suppresses the formation of competing phases, thereby yielding more uniform UTe$_2$ crystals.

\begin{figure*} \centering \includegraphics [width = 0.8\textwidth] {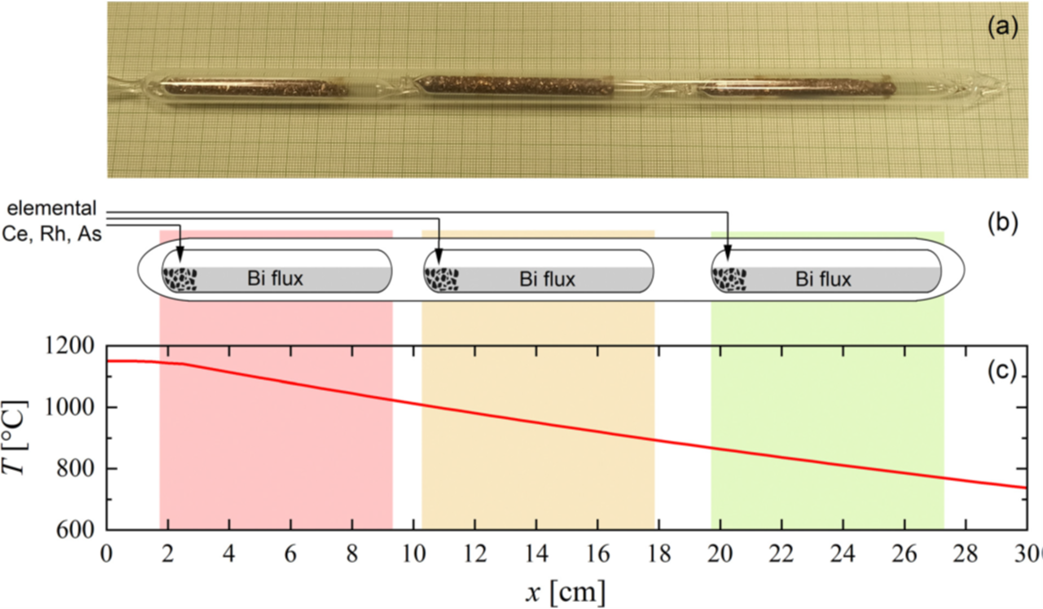}
\caption{(color online) Liquid transport growth of CeRh$_2$As$_2$ single crystals out of Sn flux. (a) Three individual ampoules are aligned in a row inside of a large ampoule. This design helps quickly determine the optimal growth temperature and/or temperature gradient. (b) Schematic picture of three separate growth units inside of the large quartz ampoule. (c) Temperature profile of the furnace. Reproduced from Ref.[\citenum{chajewski2024horizontal}] with permission from the Royal Society of Chemistry. }
\label{CeRh2As2-1}
\end{figure*}

\textit{The case of CeRh$_2$As$_2$:} CeRh$_2$As$_2$ is one of those rare materials that show multiphase superconductivity. CeRh$_2$As$_2$ single crystals can be grown out of Bi flux in the vertical configuration by cooling from 1150$\degree$C to 700$\degree$C. Characterization of the crystals shows sample dependent superconducting transition temperature and weak anomalies in, for example, specific heat associated with the superconducting phase transitions\citep{chajewski2024horizontal}. This is likely due to the existence of some homogeneity range in composition around the 1:2:2 composition in the Ce-Rh-As system. The crystallization in a wide temperature range from 1150$\degree$C to 700$\degree$C results in composition inhomogeneity responsible for the smeared anomalies in specific heat and sample dependent superconducting temperature. This motivated Chajewski \textit{et al.} to try the LTG of  CeRh$_2$As$_2$ crystals out of Bi flux.

Figure\,\ref{CeRh2As2-1} shows the growth ampoules and temperature profile for the growth. One unique feature of this work is that three separate ampoules are aligned in a row within a two-zone horizontal tube furnace with the temperature gradient spanning nearly the entire range typically utilized in a conventional vertical flux growth. This design enables simultaneous growth of three distinct samples within a single tube furnace, facilitating rapid determination of optimal temperatures for producing high quality CeRh$_2$As$_2$ single crystals. Detailed characterization of crystals obtained from different ampoules indicates that LTG yields superior quality crystals compared to those from a vertical flux growth and the highest quality crystals are achieved through LTG conducted at lower temperatures.

All three compounds presented above exhibit temperature dependent nonstoichiometry and LTG has been shown to yield better crystals than conventional vertical flux growth. The key relevant distinction between these two techniques is that crystallization in LTG occurs at a nearly constant temperature, whereas crystallization in conventional vertical flux growth proceeds during cooling over a wide temperature range. For example, conventional vertical flux growth requires cooling from 1150$\degree$C to 700$\degree$C for CeRh$_2$As$_2$, from 950$\degree$C to 650$\degree$C for UTe$_2$, and from 1200$\degree$C to 500$\degree$C for YFe$_2$Ge$_2$. These temperature intervals are much larger than the temperature variation experienced near the crystallization region in a typical LTG growth which is normally in the range of 5-10 degrees when the natural temperature gradient from the furnace is used. As a result, LTG can reduce the growth temperature driven changes in stoichiometry and defect concentration during crystal growth. When needed, the temperature profile near the crystallization zone can be deliberately engineered to minimize local temperature variation and reduce sample-to-sample variation. Conversely, the nonstoichiometry of the grown crystals can also be intentionally tuned by adjusting the crystallization temperature near the cold end. Such control is difficult to achieve in conventional vertical flux growth by varying only the starting composition. This capability may be particularly useful for chemical substitution studies when the physical properties are highly sensitive to doping level, especially if the dopant distribution coefficient varies strongly with crystallization temperature and/or very fine doping adjustments are needed.

The benefit of maintaining a nearly constant crystallization temperature may extend more broadly to materials whose defect chemistry is sensitive to the growth temperature. The examples of LuNb$_6$Sn$_6$, MoTe$_2$, and WTe$_2$ discussed below are consistent with this possibility, although the underlying microscopic mechanisms is not yet fully understood.

\textit{The case of LuNb$_6$Sn$_6$:} LuNb$_6$Sn$_6$ is a recently discovered Nb-based kagome metal belonging to the HfFe$_6$Ge$_6$-type \textit{A}\textit{M}$_6$\textit{X}$_6$ family and represents the first 4$d$-based member of this structural class \cite{Ortiz2025LnNb6Sn6}. Small, well faceted single crystals with a typical mass of 3-5\,mg can be obtained by the conventional vertical flux growth starting with a mixture of Lu:Nb:Sn=8:2:90. The limited crystal size and yield are believed to result from the low solubility and/or diffusivity of Nb in Sn flux. Although these crystals are sufficient for many measurements, they are too small for experiments requiring large sample volumes, such as neutron scattering. Motivated by this need, we explored LTG using a mixture of Lu and Sn with molar ration of Lu:Sn=1:10 as the flux. 

In a typical LTG growth, about 3\,g Nb slugs (Alfa Aesar, 1/8 inch diameter, 1/8 inch length) were first loaded into a quartz tube (19 mm outer diameter, 1.5 mm wall thickness) sealed at one end.  About 60\,g of mixture of Lu pieces (Ames Laboratory) and Sn shots (Alfa Aesar, 99.99\%) were added to fill the tube and then the tube was sealed under vacuum. The total length of the growth ampoule is about 10\,cm. The growth ampoule was placed in a single zone horizontal tube furnace with Nb slugs sitting at the hot end maintained at 950$\degree$C and the cold end at 910$\degree$C. After three weeks, the furnace was powered off. The cold end of the ingot containing the crystals was cut off, put inside of 5\,ml Canfield Crucible Set \cite{canfield2016use}, resealed in an evacuated quartz tube, and centrifuged to remove the remaining Sn flux. Near 4.5\,g of LuNb$_6$Sn$_6$ single crystals were recovered from the final 2\,cm of the ingot at the cold end. Individual crystals reached masses of up to 250 mg per crystal and dimensions of approximately 1\,cm\,$\times$\,1\,cm\,$\times$\,0.1\,cm, an increase of more than two orders of magnitude in crystal mass compared with those by the conventional vertical flux growth. Such a combination of high total yield and large individual crystal size is unlikely to be achieved using the conventional vertical flux technique. The availability of these large crystals has enabled neutron scattering studies, which are currently in progress and will be reported elsewhere. A similar growth challenge arises for V-containing \textit{A}V$_6$Sn$_6$ compounds because of the limited solubility of V in Sn flux. The same LTG strategy has also been successfully applied to ScV$_6$Sn$_6$, suggesting that LTG may be broadly useful for growing other 166 compounds whose constituent elements have limited solubility in the flux.

Interestingly, LuNb$_6$Sn$_6$ crystals grown using LTG show a structurally driven density wave at 85\,K, which is 17\,K higher than that observed in crystals grown by the conventional vertical flux technique. The detailed origin of this difference is still under investigation. One possible explanation is that crystal growth at a constant temperature in LTG affects the defect concentration, chemical homogeneity, or stoichiometry, thereby stabilizing the density wave transition at a higher temperature.

The unusually large size of LuNb$_6$Sn$_6$ crystals obtained by LTG is unexpected and provides useful insight into how crystal size may be controlled in this technique. In almost all materials grown in our laboratory, crystals obtained from LTG have dimensions comparable to those grown by conventional vertical flux growth, although LTG typically produces a much larger quantity of crystals in a single growth. The LuNb$_6$Sn$_6$ case suggests that LTG can also promote the growth of much larger individual crystals under favorable kinetic conditions. In Sn flux, the solubility of Nb is very limited, and the effective supply rate of Nb to the crystallization front may be controlled by a combination of charge dissolution at the Nb/Sn interface, the low concentration of dissolved Nb in the melt, mass transport through the flux, and incorporation into the growing crystals. Under such a dynamic condition, Nb may be supplied gradually from the hot end to the cold end over an extended period of time. This slow and continuous supply may maintain a modest supersaturation near the crystallization front, suppress excessive nucleation, and allow existing crystals to grow for a longer time. It is a reasonable guess that a rapid generation of supersaturation would promote the formation of many nuclei and distribute the available solute among many smaller crystals. Thus, the large LuNb$_6$Sn$_6$ crystals grown by LTG may result from a favorable combination of continuous solute supply, limited Nb solubility, slow effective Nb transport, and proper nucleation and growth kinetics.

This observation indicates that the kinetic conditions in LTG may be optimized not only to increase total crystal yield but also to enhance the size of individual crystals. The spatial separation between the charge and the crystallization zone provides the necessary experimental flexibility: the charge dissolution rate, mass transport, supersaturation, nucleation and growth rate can in principle be tuned by adjusting the temperature profile along the growth ampoule and by engineering the geometry of the growth ampoule as discussed later. In practice, however, such optimization is difficult because the relevant thermodynamic and kinetic data, including solubility, liquidus relations, dissolution rates, diffusivities, and possible convective transport in the melt, are often unavailable. Qualitatively, a lower dissolution temperature or a smaller temperature difference between the charge dissolution and crystallization may provide milder growth kinetics and reduce nucleation density, but these conditions may also reduce the solute supply rate and require a longer growth time. This leads to a general challenge in LTG that is discussed further in the Experimental Consideration section: the appropriate growth time is difficult to estimate before a growth is attempted because it depends on coupled processes including charge dissolution, mass transport through the melt, supersaturation, nucleation and growth. To our knowledge, these kinetic aspects of LTG have not yet been systematically investigated, and a quantitative understanding of them would be valuable for designing growth recipes with better control over crystal size, yield, and quality.

\textit{The case of MoTe$_2$ and WTe$_2$:} The recent realization of dissipationless fractional Chern insulating states in twisted MoTe$_2$ further highlights the importance of the quality of starting crystals used for the devices \cite{Park2026FCI}. The residual longitudinal dissipation coexisting with quantized transverse resistance in earlier studies may be associated with imperfect sample quality. This motivated significant efforts to reduce lattice disorder in the bulk MoTe$_2$ crystals used to fabricate twisted bilayer devices. To reduce the defect density, Park et al. explored LTG of MoTe$_2$ crystals using Te as the self flux. Approximately 40 g of Mo and Te with a molar ratio of 1:20 were sealed in a quartz ampoule equipped with a narrow neck between the charge and growth regions. After keeping the hot end at 600$\degree$C while the cold end was held at 550$\degree$C for about one month, the growth ampoule is tilted to separate crystals from flux with the neck keeping large crystals in the growth region. Excess Te flux on crystals was removed by annealing the crystals overnight at 380$\degree$C in another evacuated quartz tube. Conductive atomic force microscopy revealed a defect density of approximately 2$\times$10$^9$cm$^{-2}$, about two orders of magnitude lower than that of commercially available MoTe$_2$ crystals. These high quality crystals enabled the fabrication of twisted bilayer MoTe$_2$ devices exhibiting significantly reduced moiré disorder, quantized anomalous Hall resistance together with vanishing longitudinal resistance at the fractional filling, and the first observation of a dissipationless fractional Chern insulating state.

Delgado et al. applied the same growth strategy to grow WTe$_2$ crystals with residual resistivity ratios as high as 12200 and transport mobilities up to 7$\times$10$^5$cm$^2$V$^{-1}$s$^{-1}$, far exceeding previous reports \cite{Delgado2026WTe2}. Scanning tunneling microscopy further revealed an extremely low surface defect density of approximately 2.9$\times$10$^{-4}$ per unit cell. Together, the MoTe$_2$ and WTe$_2$ studies illustrate that LTG can substantially suppress crystal defects in transition metal dichalcogenides and enable the discovery and investigation of quantum phenomena that are otherwise obscured by disorder.

As in the previous examples, the ability of LTG to maintain crystallization at a relatively low and nearly constant temperature likely plays a key role in producing crystals with exceptionally low defect densities. The narrow neck separating the charge dissolution and crystallization may also be important although its role remains to be answered. Practically, the narrow neck facilitates separating crystals from remaining flux by allowing the growth ampoule to tilt after growth. In addition, the reduced cross section near the neck may also influence mass transport through the melt by suppressing convection and thus regulating the supply of solute to the growth front. This is discussed further in the Experimental Consideration section. This neck design inspired us to prepare similar but long neck growth ampoules that could help reduce the amount of expensive flux required while still allowing efficient liquid transport. Motivated by the exceptionally low defect densities observed in MoTe$_2$ and WTe$_2$, an important open question is whether geometrical control of mass transport can be used more generally to suppress defect formation and improve crystal quality across a broader range of materials grown by LTG.

\subsection{Experimental considerations}

\textit{Furnace selection:} Both single-zone and two-zone tube furnaces have been used in our laboratory for LTGs, provided that an appropriate temperature gradient is established along the growth ampoule. We have routinely employed a single-zone Thermo Scientific Mini-Mite tube furnace with a maximum temperature of 1100$\degree$C. Its 12-inch heating zone allows us to perform two growths simultaneously. The natural temperature gradient depends strongly on the thermal insulation placed at the ends of the furnace. With insulating plugs, the maximum temperature difference along a typical growth ampoule of 4 inch is about 20$\degree$C. Without insulation, the temperature gradient can be increased to approximately 40$\degree$C. Extending the growth ampoule out of the heating zone can provide an even larger temperature difference between the hot end and the cold end. However, this is not recommended as the temperature gradient near the cold end can be very large. The single zone tube furnace has only one thermocouple located at the center of the furnace. We always put an external thermocouple at the cold end of each growth ampoule to monitor its temperature during crystal growth.

A two-zone furnace with the maximum temperature of 1200$\degree$C offers greater flexibility because the magnitude of the temperature difference and sometimes the temperature profile (dT/dx) along the ampoule can be adjusted independently. This provides a better optimization for different materials. To evaluate the practical importance of furnace design, we carried out LTG of MoTe$_2$ using both single-zone and two-zone furnaces under comparable growth conditions. No significant difference in crystal quality or growth behavior was observed. These results suggest that a single-zone tube furnace is sufficient and provides a simple and cost effective solution for most materials or initial proof-of-principle test growths. A two-zone furnace becomes advantageous when a larger temperature difference or a more carefully engineered temperature profile is required for the growth.

Regardless of whether a single-zone or two-zone furnace is used, we usually first characterize the temperature profile along an empty growth ampoule at several furnace set temperatures before crystal growth. Because the presence of molten flux can alter heat transfer and thereby modify both the temperature difference and the local temperature gradient along the ampoule, we occasionally repeat the calibration using an ampoule containing a similar amount of the intended flux when such measurements are necessary and can be performed safely. The temperature calibration is important because the absolute temperature and temperature profile along the ampoule play a central role in determining dissolution, mass transport, nucleation, and phase selection during growth.

\textit{Melt stability:} One of the key advantages of LTG is that charge dissolution and crystal precipitation occur at opposite ends of the growth ampoule and are therefore spatially separated. The supersaturation required for crystal growth at the cold end is generated continuously by dissolution of the charge at the hot end and subsequent transport of the dissolved species through the molten flux. Consequently, a successful LTG requires the molten flux to continuously transport the dissolved charge from the dissolution region to the crystallization region. However, this transport condition can break down when the charge is dissolved into the molten flux faster than it can be transported to the cold end and incorporated into the growing crystals. Under these conditions, the concentration of dissolved species in the melt increases and may alter its thermophysical properties, including wetting behavior with the growth ampoule, viscosity, density, and surface tension. In the worst case, the molten flux contracts into one or several droplets instead of remaining distributed along the growth ampoule, thereby terminating effective liquid transport.

We encountered this problem during some test growths of i-Sc$_{12}$Zn$_{88}$ quasicrystals and 2H-MoTe$_2$ single crystals using LTG. For i-Sc$_{12}$Zn$_{88}$, at a growth temperature of 450~$^\circ$C, the Zn melt contracted into a single droplet due to the high surface tension when the Sc/Zn molar ratio exceeded approximately 0.3. For MoTe$_2$, liquid transport occasionally failed when reduced Mo powder was used as the charge, whereas such failures were rarely observed when Mo shots were used. The much larger surface area of the powder likely accelerates dissolution of Mo into the Te melt, which breaks the dynamic balance between charge dissolution, mass transport, and crystal precipitation. Possible remedies include reducing the exposed surface area and starting amount of the charge, lowering the hot-end temperature to moderate the dissolution rate, increasing the amount of flux to dilute the dissolved charge, or adjusting the temperature profile and ampoule geometry to modulate transport of dissolved charge toward the crystallization region. The effectiveness of these approaches is expected to be materials dependent because they influence charge dissolution, transport, wetting, convection, and crystallization simultaneously.

Those failed growths of i-Sc$_{12}$Zn$_{88}$ quasicrystals and 2H-MoTe$_2$ suggest that the stability of LTG is governed by the dynamic balance among charge dissolution, mass transport through the molten flux, and crystal precipitation, rather than by the initial charge/flux ratio or the temperature profile alone. A quantitative understanding of these coupled processes remains lacking and deserves systematic experimental and theoretical investigation.

\textit{Growth time:} Unlike conventional vertical flux growth, where the growth time is often tied to the cooling schedule, LTG can be performed under a fixed temperature profile for a chosen amount of time. Growth time is one important but largely empirical parameter in LTG at this stage. In many of our exploratory growths, we typically maintain the ampoule at the growth temperature for 2-4 weeks. This choice is not based on a quantitative estimate of the growth rate, but in practice it has been sufficient to obtain enough quantity of crystals for our planned measurements. Such an empirical approach is often adequate when the main goal is to obtain usable crystals rather than to optimize crystal size, yield, or defect concentration. In principle, however, the appropriate growth time should depend on the coupled kinetics of charge dissolution, mass transport through the melt, supersaturation, nucleation, and crystal growth. These processes are affected by the temperature profile, charge/flux ratio, rate of charge dissolution, solubility of the charge in the flux, thermophysical properties of the melt, ampoule dimensions, filling fraction, and ampoule geometry. A longer growth time pushes the growth front toward the hot end and increases the total crystal yield. This has been confirmed in our growths of Mo$_3$Sb$_7$, MoTe$_2$, and Fe$_3$Sn$_2$. In several 4-week long growths of Mo$_3$Sb$_7$ and MoTe$_2$, we found crystals throughout the whole ampoule.  Future systematic time dependent growth studies would be valuable for understanding the growth kinetics and for improving control over crystal yield and quality.

\begin{figure} \centering \includegraphics [width = 0.48\textwidth] {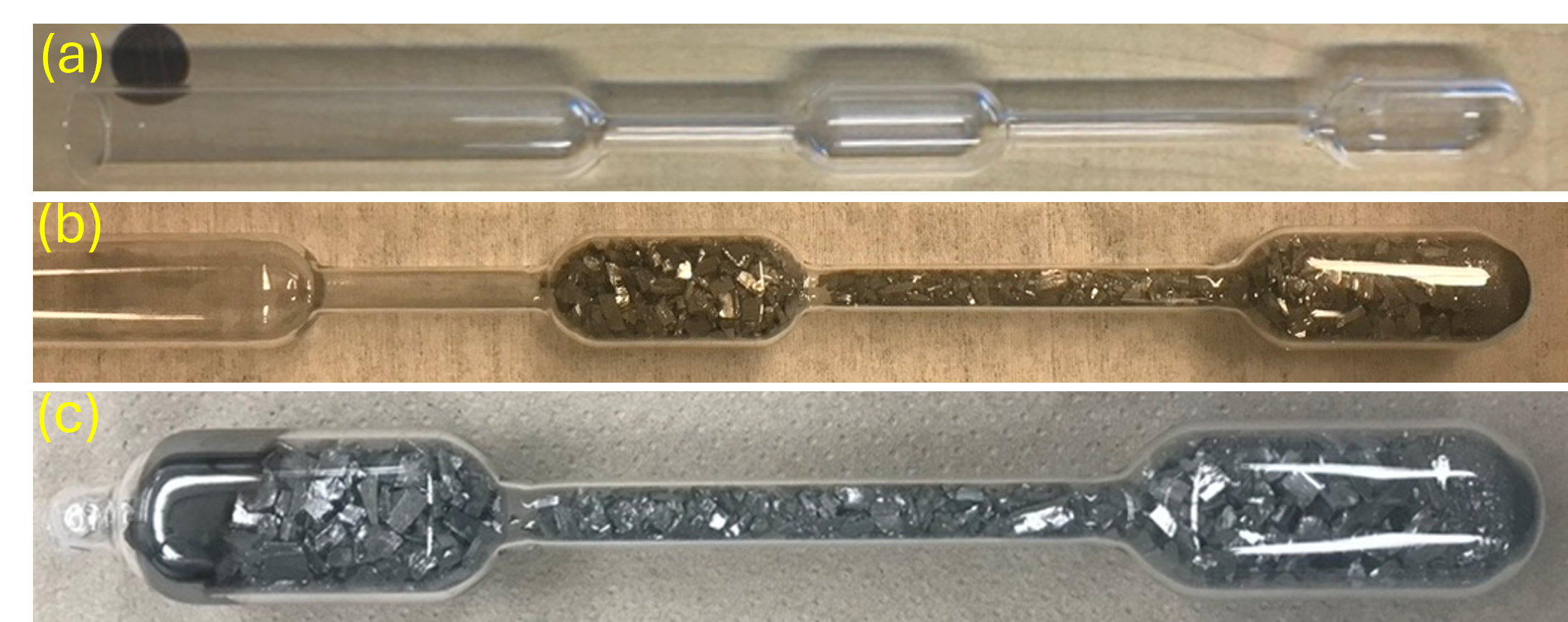}
\caption{(color online) Dumbbell shaped quartz ampoule designed for the liquid transport growths. (a) An as-made quartz ampoule. (b) Filled with Te flux and Mo charge. (3) A sealed ampoule ready for the growth. The ampoule was made by Carlos Flores and Jason Craig at ORNL glass shop.}
\label{ampoule-1}
\end{figure}

\textit{Growth ampoule:} For a LTG performed in a single zone tube furnace, the length of the growth ampoule depends on the natural temperature gradient of the furnace at the setup temperature and the required temperature gradient for the crystal growth. The growth ampoule can be rather long when a two-zone tube furnace is used. A long ampoule requires a large amount of flux, which can be costly. One way to reduce the amount of flux is to use a constricted or dumbbell-shaped quartz ampoule, as shown in Fig.~\ref{ampoule-1}. Constricted horizontal ampoules have been used historically in molten salt solution growth. For example, Johnson and Parker employed a constricted quartz tube for the growth of cubic ZnS from molten KCl and noted that the constriction helps control material transfer and separate the raw material and crystal growth regions \cite{Johnson1968ZnS}. Motivated by this historical example and by recent narrow neck ampoule designs used for MoTe$_2$ growth \cite{Park2026FCI}, we tested a dumbbell shaped quartz ampoule for LTG. The long slim neck in the design allows mass transport through the melt and reduces the amount of flux needed. The neck can also help separate crystals from flux by tilting the growth ampoule if the obtained crystals are larger than the inner diameter of the neck. To make a long, uniform neck as shown in Fig.\,\ref{ampoule-1} is time consuming and requires additional glassblowing skill. Often we simply heat the quartz ampoule when it is being pumped to collapse part of the quartz ampoule to reduce the dimension as shown in Fig.\,\ref{CrPD-1}.

One concern with the dumbbell-shaped ampoule design is whether the narrow neck, which is filled with molten flux during the growth, could generate sufficient stress to crack the quartz tube when the flux expands upon solidification. This is particularly relevant for fluxes such as Bi, which undergo a volume expansion on freezing. To evaluate this possibility, we tested the ampoule design in LTG of MnBi using Bi flux. Among four growth ampoules examined, two developed cracks in the quartz after cooling. Importantly, however, the cracks did not compromise the integrity of the sealed ampoules. No leakage of molten flux was observed, and the materials inside remained free of oxidation, indicating that the cracks likely formed only after the flux had completely solidified. These results suggest that although fluxes exhibiting expansion upon solidification increase the risk of quartz cracking, the dumbbell shaped ampoule remains a practical and generally safe design for LTG. 

An important open question is how the dumbbell shaped ampoule, particularly its long narrow neck, influences mass transport during LTG. More broadly, the dominant mass transport mechanism in a regular cylindrical LTG ampoule without a narrow neck is also not yet well understood. In a horizontal ampoule with an imposed temperature gradient,  mass transport may involve both diffusion and convection. Thermal gradients, composition gradients produced by dissolution of the charge, and surface tension gradients at the melt surface can all drive convective flow in the molten flux. The relative importance of diffusion and convection should depend on the melt properties, temperature gradient, ampoule diameter, filling fraction, wetting behavior, and geometry of the growth ampoule. The thermophysical properties of the melt that should be considered include, but are not limited to, viscosity, density, solute diffusivity, surface tension.  The long narrow neck of the dumbbell shaped ampoule is expected to suppress convective circulation and may make transport more diffusion limited than in a regular cylindrical ampoule. This raises an interesting possibility for the high quality MoTe$_2$ and WTe$_2$ crystals grown using narrow neck ampoules: in addition to enabling growth at a low and nearly constant temperature, the ampoule geometry may also help reduce convective fluctuations in solute supply and thereby contribute to lower defect densities. This possibility remains speculative, but it deserves systematic investigation. A quantitative treatment of mass transport in LTG is beyond the scope of the present work. Future studies combining controlled growth experiments with different ampoule geometries and temperature profiles together with theoretical modeling of diffusion and convection would be valuable for establishing a more predictive understanding of LTG.

\textit{When LTG may not be advantageous:} Although the examples discussed in this work demonstrate several distinctive advantages of LTG, it should be regarded as a complementary crystal growth technique rather than a universal replacement for conventional vertical flux growth. Even when thermodynamic considerations suggest that LTG should be feasible, kinetic or practical limitations may prevent successful growth. Our previous LTG of MnBi provides one such example: at temperatures where MnBi precipitates, slow solute diffusion and the relatively high viscosity of the Bi-rich melt may limit mass transport, and we generally obtained small crystals containing Bi flux inclusions \cite{yan2017flux}. More generally, any factor that disrupts the dynamic balance among charge dissolution, mass transport, and crystallization can destabilize the growth. Systems involving high vapor pressures or chemical incompatibility with quartz or other crucible materials also pose practical and safety concerns. We have encountered quartz tube failures in some experiments using salt fluxes as the transport agent. Multicomponent compounds present an additional challenge when the constituent elements differ substantially in solubility or diffusivity, particularly when competing phases are stable under similar conditions. From our current understanding, LTG is most useful when spatial separation of charge dissolution and crystallization directly addresses the primary limitation of conventional flux growth for a particular material.

\subsection{Summary and outlook}

In summary, we reviewed recent progress on LTG as a horizontal flux growth technique that complements the conventional vertical flux growth for producing high quality single crystals of quantum materials. Early examples of salt flux transport growths under a temperature gradient were reported in the late 1960s, while work in recent years has generalized this concept into a broadly useful horizontal flux growth technique. The unique capabilities of LTG can be traced to its three key characteristics: charge dissolution and crystallization are spatially separated; dissolution and crystallization occur at different temperatures; and crystallization can proceed at a nearly constant temperature. These characteristics distinguish LTG from conventional vertical flux growth and make it particularly useful in the following three situations: 

(1) \textit{When a large quantity of crystals is needed}. Because crystallization begins while undissolved charge remains in the source region, the charge dissolution continuously supplies the dissolved solute consumed during crystal growth at the cold end. The total amount of charge used can therefore greatly exceed the amount soluble in the flux at any one time, so the attainable yield is not directly limited by the equilibrium solubility of the charge.  This advantage is particularly valuable when one or more constituents have limited solubility or when large sample volumes are required. The growths of LuNb$_6$Sn$_6$  and YFe$_2$Ge$_2$ further suggest that, under favorable kinetic conditions, LTG can also produce unusually large individual crystals. 

(2) \textit{When the desired phase crystallizes only within a narrow temperature and/or composition window}. LTG separates charge dissolution from crystallization, allowing the hot end to be optimized for charge dissolution while the cold end temperature determines the crystallizing phase. This greatly relaxes the stringent constraints encountered in conventional vertical flux growth, as demonstrated by Fe$_3$Sn$_2$ and CrTe$_3$.

(3) \textit{When the stoichiometry, defect concentration, and thus physical properties of the target material are sensitive to the crystallization temperature}. By allowing crystallization to proceed at a low and nearly constant temperature, LTG improves chemical homogeneity and crystal quality in a wide variety of compounds. Recent examples such as YFe$_2$Ge$_2$, UTe$_2$, CeRh$_2$As$_2$, MoTe$_2$, WTe$_2$, and LuNb$_6$Sn$_6$ illustrate how LTG can improve crystal quality when physical properties are highly sensitive to growth-temperature-dependent nonstoichiometry or disorder.

Despite these useful applications, the practical implementation of LTG remains largely empirical at this stage. In this review, we discussed several experimental considerations, including furnace selection, melt stability, growth time, and geometry of growth ampoule. A single-zone tube furnace is often sufficient for exploratory growths, whereas a two-zone furnace is useful when a larger temperature difference or a more carefully engineered temperature profile is required. A stable LTG requires an appropriate dynamic balance among charge dissolution, mass transport, and crystal precipitation. At present, however, the coupled kinetics governing these processes remain poorly understood. Growth time is another important but poorly understood parameter. In many exploratory growths, dwell times of 2-4 weeks are sufficient to obtain useful crystals, but the optimal time should depend on the coupled kinetics of charge dissolution, mass transport, supersaturation, nucleation, and crystal growth. Ampoule geometry also plays an important role. Dumbbell-shaped or narrow-neck ampoules can reduce the amount of flux required and facilitate crystal/flux separation, but they may also modify mass transport by suppressing convective circulation and regulating the supply of dissolved species to the crystallization region.

Looking forward, a major opportunity is to transform LTG from an empirical growth method into a more predictive and data assisted crystal growth technique.  At present, LTG experiments are optimized through trial and error due to limited information about charge dissolution, solute transport, nucleation, and growth-front evolution inside growth ampoules.  These information is particularly important because the effective transport rate through the melt controls supersaturation, nucleation density, crystal size, yield, and defect concentration. In situ and time-resolved studies, for example using synchrotron X-ray and neutron diffraction and imaging, thermal imaging, or interrupted growth experiments combined with composition mapping, could provide valuable information about phase formation, mass transport, supersaturation, and growth kinetics under realistic temperature gradient conditions. Future studies combining controlled growth experiments with different temperature profiles, ampoule dimensions, filling fractions, neck geometries, and growth times would be valuable and the outcomes could enable machine learning or Bayesian optimization approaches to identify useful trends. Combining in situ diagnostics, thermodynamic and transport modeling, and data-driven optimization may help establish LTG as a more predictive platform for growing high quality quantum materials

An equally exciting opportunity lies in exploiting the unique design flexibility offered by this growth technique. LTG allows the dissolution region, transport pathway, and crystallization region to be engineered separately. This opens new possibilities for tailoring crystal composition, defect density, dopant incorporation, nucleation density, and growth kinetics. With improved understanding and control, LTG should become an increasingly useful technique for the synthesis of high quality single crystals and for the discovery and investigation of emerging quantum materials.

\section{Acknowledgment}
We thank Dr. Chaowei Hu for helpful discussions. This work was supported by the US Department of Energy, Office of Science, Basic Energy Sciences, Materials Sciences and Engineering Division. The growth of LuNb$_6$Sn$_6$ was sponsored by the Laboratory Directed Research and Development Program of Oak Ridge National Laboratory, managed by UT-Battelle, LLC, for the US Department of Energy.

\section{references}
%

\end{document}